%% file: Arxiv_Submission.tex
\documentclass[]{fairmeta}

\usepackage{xcolor}
\usepackage{subcaption}
\usepackage{graphicx}
\usepackage{svg}

\newcommand{\prodecasa}[1]{{PDCaSA}}

\title{Proactive Detection and Calibration of Seasonal Advertisements with Multimodal Large Language Models}

\author[1 ,\dagger]{Hamid Eghbalzadeh}
\author[1 \dagger]{Shuai Shao}
\author[1 \dagger]{Saurabh Verma }
\author[1 \dagger]{Venugopal Mani}
\author[1 \dagger]{Hongnan Wang}
\author[1 ]{Jigar Madia}
\author[1 ]{Sean Obyrne}
\author[1 \dagger]{Vitali Karpinchyk}
\author[1 ]{Andrey Malevich}

\affiliation[1]{Meta, Menlo Park, USA}

\contribution[\dagger]{These authors contributed equally
\\
\{heghbalz,sshao,saurabh08,venugopalmani,hongnanwang,jmadia,seanobyrne,vitalikk,amalevich\}@meta.com}
\abstract{A myriad of factors affect large scale ads delivery systems and influence both user experience and revenue. 
One such factor is proactive detection and calibration of seasonal advertisements to help with increasing conversion and user satisfaction. 
In this paper, we present \textbf{P}roactive \textbf{D}etection and \textbf{Ca}libration of \textbf{S}easonal \textbf{A}dvertisements (\textbf{\prodecasa}), 
a research problem that is of interest for the ads ranking and recommendation community, both in the industrial setting as well as in research.
Our paper provides detailed guidelines from various angles of this problem tested in, and motivated by a large-scale industrial ads ranking system.
We share our findings including the clear statement of the problem and its motivation rooted in real-world systems, evaluation metrics, and sheds lights to the existing challenges, lessons learned, and best practices of data annotation and machine learning modeling to tackle this problem.
Lastly, we present a conclusive solution we took during this research exploration: to detect seasonality, we leveraged Multimodal LLMs (MLMs) which on our in-house benchmark achieved 0.97 top F1 score.
Based on our findings, we envision MLMs as a teacher for knowledge distillation, a machine labeler, and a part of the ensembled and tiered seasonality detection system, which can empower ads ranking systems with enriched seasonal information.}

\date{\today}

\begin{document}
\maketitle

\section{Introduction}
\label{sec:intro}
\input{sections/intro}

\section{Motivation}
\label{sec:motivation}
\input{sections/motivation}

\section{Ground Truth Collection}
\label{sec:ground_truth_collection}

\input{sections/gt_collection}

\section{PDCaSa}
\label{sec:method}
This section details the task of Proactive Detection and Calibration of Seasonal Advertisements (\prodecasa{}).
\input{sections/method}

\section{Experimental Results}
\label{sec:experimental_setup}
\input{sections/experimental_setup}

\section{Conclusion}
Being able to understand, detect, model, and treat seasonality is important for ads delivery. 
In this work, we have utilized MLMs to detect whether an ad is seasonal. 
We finetuned LLAVA and CLIP models with classification heads which resulted in a top 0.97 F1 score. 
Additionally, we shared our learned lessons and best practices for the community, along with a large set of ablation results that sheds light on the challenges and difficulty of this task, as well as best performing models for various usecases such as improving labeling pipelines and scalable efficient models.

We envision MLM as a teacher for knowledge distillation, a machine labeler, and a part of the ensembled and tiered seasonality detection system. 

\clearpage
\newpage
\beginappendix
\section{Related Work}
\label{sec:related_work}
\input{sections/related_work}

\section{MLM Model ablations}
\label{sec:ablations}
\begin{figure}[!t]
\centering
\begin{subfigure}{0.42\textwidth} 
    \centering
    \includegraphics[width=\textwidth]{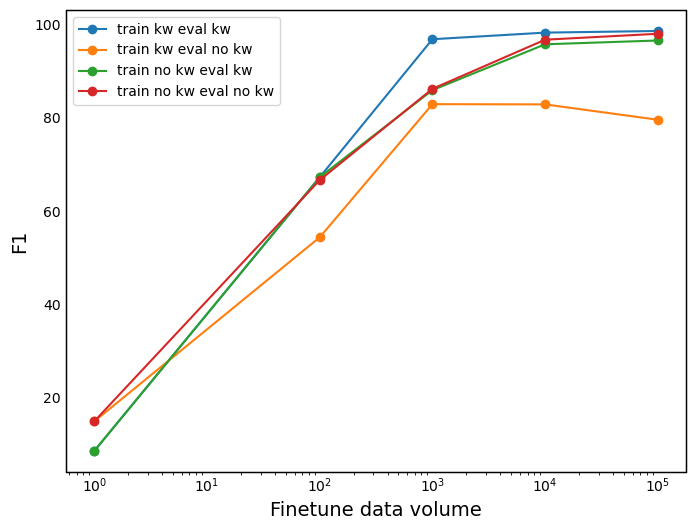}
    \caption{Finetune volume (e2e MLM)}
    \label{fig:volume}
\end{subfigure}\hfill
\begin{subfigure}{0.42\textwidth}
    \centering
    \includegraphics[width=\textwidth]{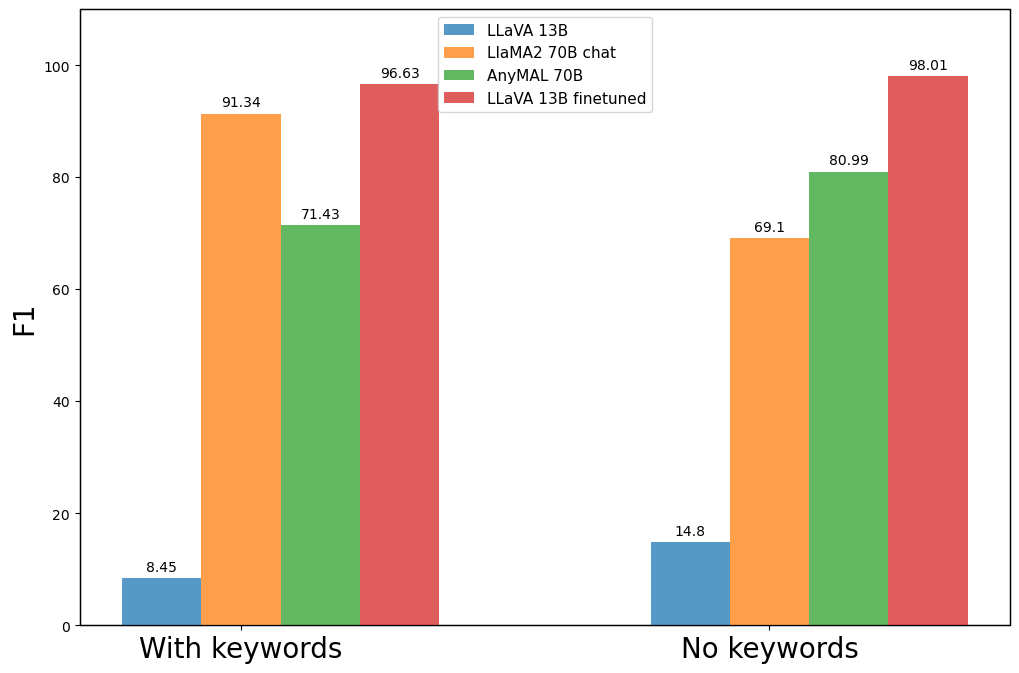}
    \caption{Model comparison (e2e MLM)}
    \label{fig:comparison1}
\end{subfigure}\\
\begin{subfigure}{0.42\textwidth} 
    \centering
    \includegraphics[width=\textwidth]{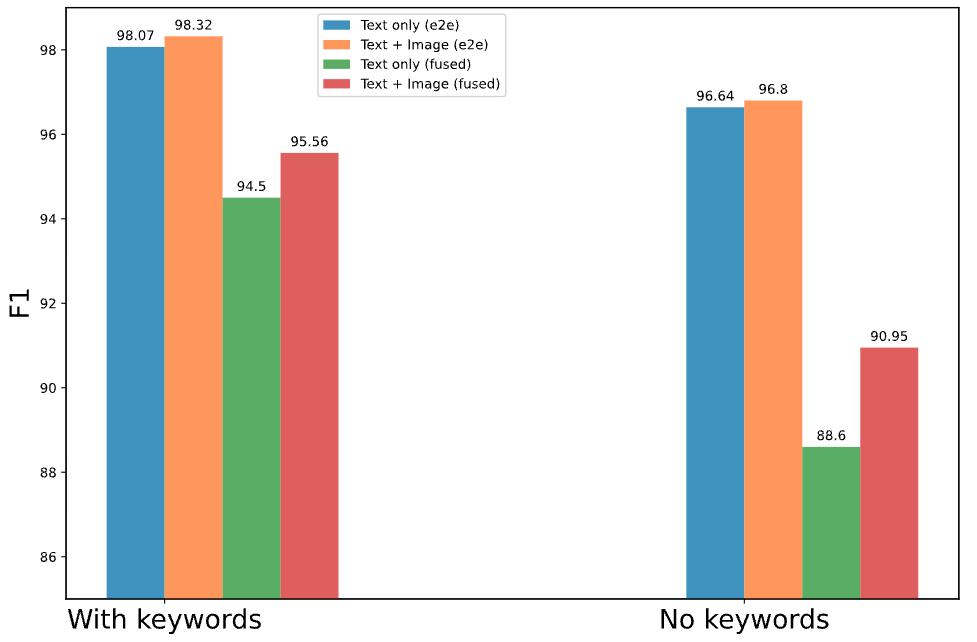}
    \caption{Modality comparisons}
    \label{fig:modality}
\end{subfigure}\hfill
\begin{subfigure}{0.42\textwidth}
    \centering
    \includegraphics[width=\textwidth]{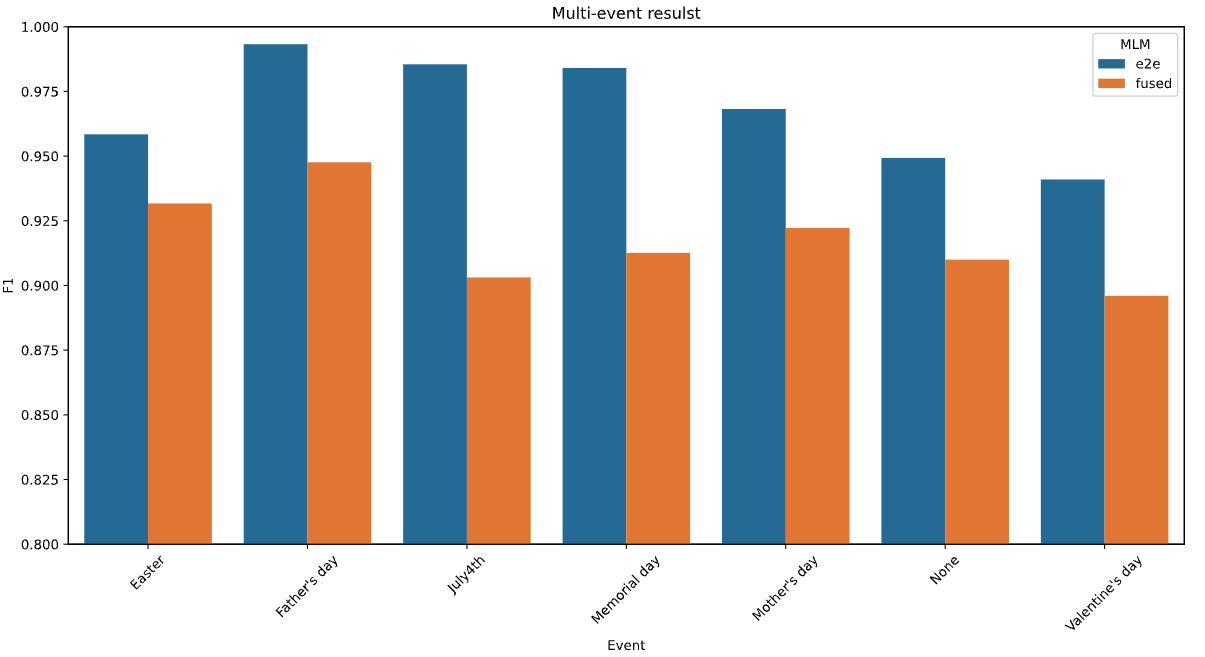}
    \caption{Multi-event results}
    \label{fig:multievent_f1}
\end{subfigure}
\caption{
(a) Model performance improves with finetune volume  
(b) An overview of model performance, evaluated by F1 score for single-event e2e MLM. 
(c) Incorporating additional modality like image improves the model's performance compared with text input only, validating the importance of multimodality.
d) Multievent results
}
\label{fig:three-graphs}
\end{figure}
\begin{figure}[h]
  \centering
  \includegraphics[width=0.7\linewidth]{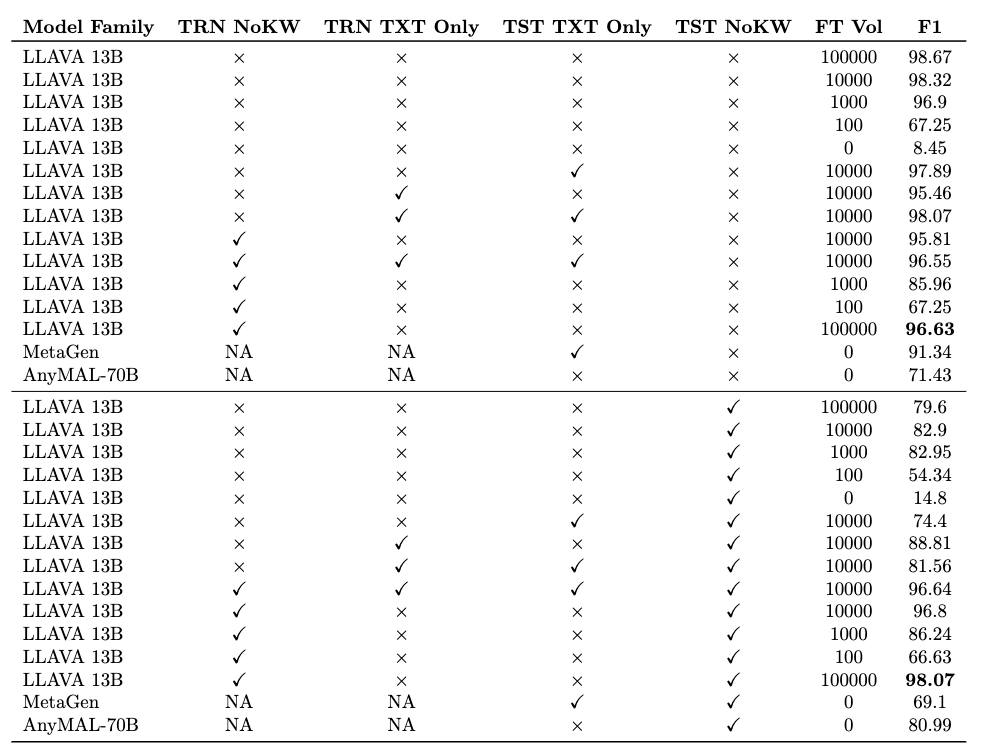}
  \caption{MLM ablations. }
\label{fig:model_perf_ablation}
\end{figure}
\textbf{Volume}:
We observed more finetune data helps with model performance, but usually ~1k to ~10k would be sufficient to bring about the best performance, depending on the task.
An interesting difference between training on texts with keywords and texts without keywords is that the performance of the former one tends to saturate fast wrt to data volume. And more finetune data can even lead to overfitting to the task and performance degradation on the harder task of predicting without keywords in texts (red line). In other words, when the model learns to recognize keywords such as “valentine’s” instead of relying on a thorough understanding of the content, it doesn't need a lot of examples.
On the other hand, on the harder task of training without keywords, more finetune data leads to better performance up to 100K (green and red line).
Another observation is the untuned model’s performance is significantly worse than the model tuned with  just a minimum amount of data (e.g. 100). This is also due to the fact that the untuned model was performing a generation task while the tuned model was performing a classification task.

\textbf{Generalization}:
Related to above, when trained on a harder task with the keywords removed in the input texts, the model is pushed to understand the whole Ad better. As a result, its capability to detect seasonality generalizes better to the eval sets with or without keywords in them.
In comparison, if we give the model an easier task, it learns to recognize or memorize the keywords, and as a result, is not able to perform well on the eval set absent of keywords (the orange curve).

\textbf{Modality}:
By comparing results with or without images in training and eval samples, we found that additional information from images helped the model to detect seasonal Ads better.
In both tasks with or without keywords in the eval or training set, adding images improved the already good performance even further.
Another interesting ablation study would be looking at only images as the input without text information. We leave that to future studies.

\section{Model Architecture}
\label{sec:model_arch}
\begin{figure}[!h]
  \centering
  \includegraphics[width=\linewidth]{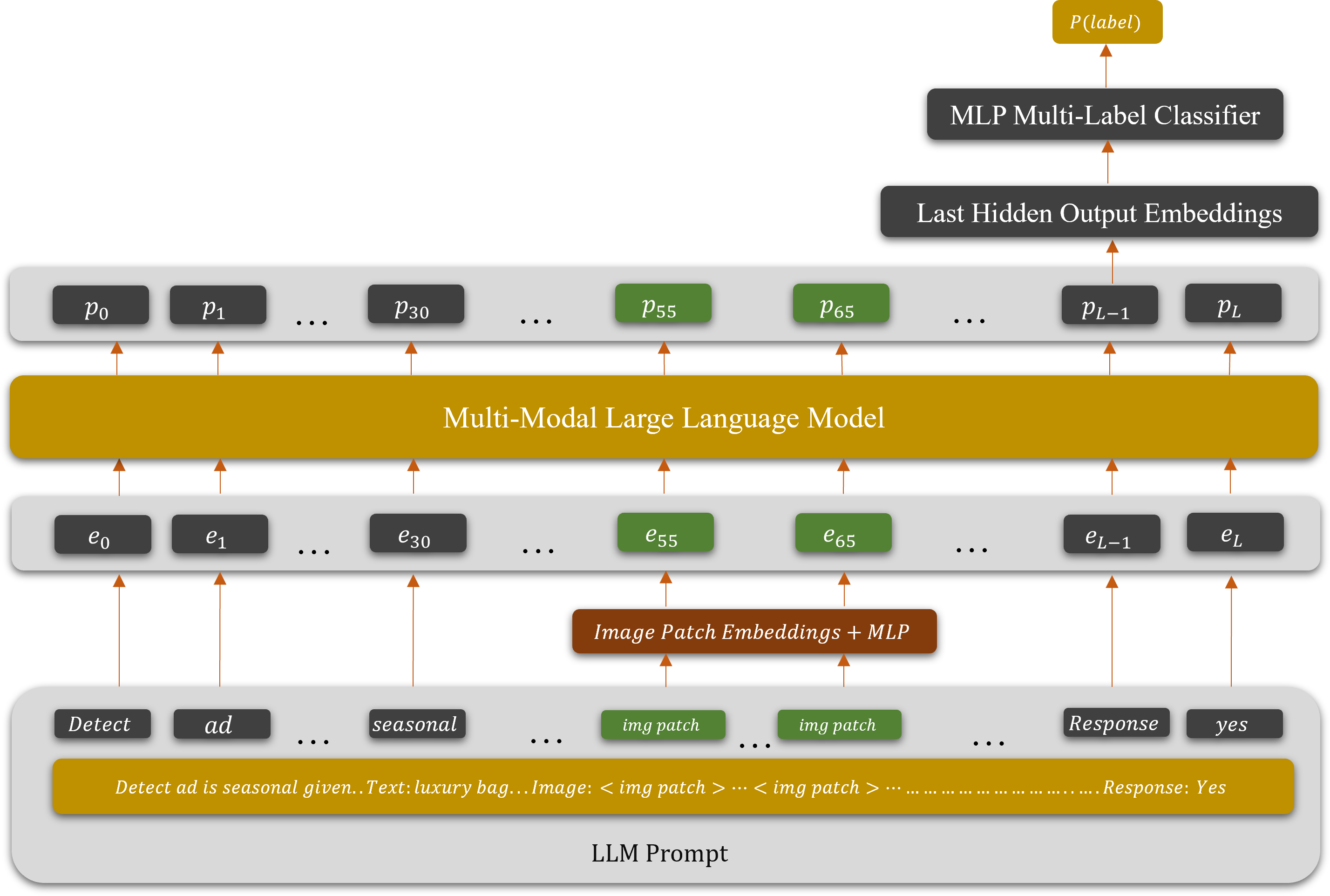}
  \caption{ We developed our model utilizing the foundational structure of the LLAVA model. In this design, the tokens for images are transformed into a continuous token embedding space and then merged with the remaining token embeddings. Consequently, the model is capable of processing image features in their respective contexts. To build the Multi-Modal LLMs Classifier as a classic Multi-Label or Multi-Class Classifier, we apply a MLP layer on the top of the LLM last hidden output layer. This layer essentially captures the information in prompt context and is responsible for generating responses in autoregressive fashion. }
\end{figure}

\clearpage
\newpage
\bibliographystyle{assets/plainnat}
\bibliography{Arxiv_Submission}

\clearpage
\newpage

\end{document}

%% file: sections/intro.tex
A \emph{seasonal ad} is a type of advertisement that is specifically designed and scheduled to coincide with a particular season, event, or holiday~\cite{stl}.
These ads are tailored to a specific time of year and the associated consumer needs, and aim to capitalize on the consumer behavior associated with these periods. For example, a toy company might run a special Ad campaign around Christmas, featuring their latest products as perfect holiday gifts for children.
Failing to differentiate and treat seasonal ads leads to missing opportunities on capitalization, and lowered user or advertiser happiness, which needs to be addressed by understanding, detecting, and treating seasonal Ads. 
The ability to proactively detect seasonal advertisements
is an important and interesting problem that serves as one of the foundations for seasonal ads treatment. 
In this work, we formulate an applied machine learning problem namely the \textbf{P}roactive \textbf{D}etection and \textbf{Ca}libration of \textbf{S}easonal \textbf{A}dvertisements (\textbf{\prodecasa}), as a research problem that is of interest for the ads ranking and recommendation community, focused around answering the question of \emph{``whether an ad is seasonal''} in an accurate, timely, and scalable manner.

Such an ads seasonality detection model can then be leveraged to identify which ads are seasonal, and which ads are not, and to further integrate seasonal information into retrieval and ranking pipelines to address the aforementioned problem. 
Though this problem encompasses various challenges:

\textbf{Determining seasonality and its categories using available data at scale}: Ads often have two main sources of information: 1) ad content and 2) interaction data. 
Ad content includes information such as text, image, videos, etc. of an ad, while interaction encompasses user interactions throughout the lifecycle of an ad, but can also include advertiser side information such as ad budget, campaign group, and so on. 
As stated before, the main goal is to develop a machine learning model that can detect whether an ad belongs to a seasonal event, or is not seasonal at all. 
Our assumption is that knowing the probability of an ad belonging to a given seasonal event will help to remedy the possible underperforming issues that may occur for the seasonal events~\cite{Yang2021}. 
Developing an ML pipeline is another contribution of our work, and to demonstrates what constraints we may face, for instance, in data annotation, modeling, and evaluation, in terms of the scale of seasonal events, number of labels, model parameters and performance, and the automation of the pipeline.

\textbf{Inherent noise}: Depending on the ad, there is always a level of uncertainty on whether an ad belongs to a certain seasonal event or not. Hence, we explore various approaches to simulate this problem in our work.
\\
\textbf{Lack of Ground truth}: Since the Ads are placed by the advertisers at large on ads platforms, in a realistic scenario they are not inherently labeled based on seasonal event, and there is no source of truth in order to identify the correct seasonal event for a given ad. 
We will elaborate more on this in the Ground Truth Collection section (Section~\ref{sec:ground_truth_collection}) and discuss the challenges it introduces.
\\
\textbf{Open-endedness}: There is some level of open-endedness to the seasonal events as there is no fixed set we can treat as an oracle. The seasonal events may change over time, and there is no guarantee that we can capture all of them all year through.
\\
\textbf{Repeatedness}: Some seasonal events (such as New Year's Eve) repeat every year at the same date, while others (such as Chinese New Year or Diwali) may change.

In order to concentrate on the ML modeling work, we focus on the two primary issues, namely, 1) determining seasonality and its category using available data, and 2) dealing with the inherent noise, while making reasonable assumptions on the rest of the open questions.
In short, the contributions of this work are as follows:
\begin{enumerate}
    \item Present the challenges of this task in an industrial-scale ranking system with billion-scale daily users.
    \item Provide scalable solutions for the identified challenges: we present a solution based on multimodal LLMs that can be leveraged in various stages of ads seasonality detection pipeline to improve performance, efficiency of the system by using the multimodal nature of ads.
    \item Demonstrate the potential of various ML paradigms and their effectiveness for this task
    \item Assess their robustness to the nature of inherent noise in ads w.r.t seasonality, by leveraging a content-based multimodal system.    
\end{enumerate}

%% file: sections/motivation.tex
Prior analyses have indicated that some ad ranking models underperform at specific times, strongly correlating with the onset of seasonal events (we provide a detailed related work in Appendix~\ref{sec:related_work}).

\begin{figure}[h]
  \centering
  \includegraphics[width=\linewidth]{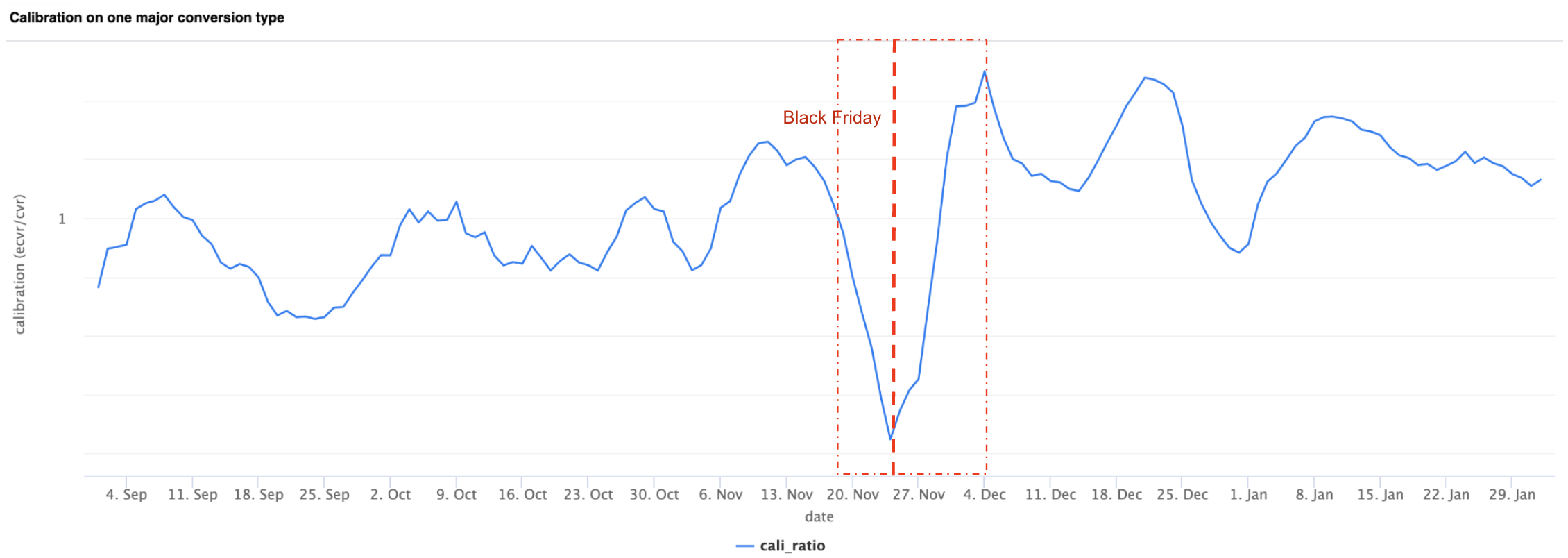}
  \caption{\label{fig:miscali} Smoothed Calibration for ads delivered on a major conversion traffic}
\end{figure}
In the Figure \ref{fig:miscali}, you can see an example of this phenomenon, visualized by the calibration plot throughout the year.
In this context, \emph{calibration} refers to the ratio of the predicted conversion rate to the actual conversion rate, with a calibration of 1 representing the ideal case.

In this example, we can observe that during the big shopping holiday "Thanksgiving-Black Friday" period, the calibration is much lower than 1, that can be interpreted as the under-calibration of a ranking system.

Based on such evidences, we define the \emph{seasonal event} as a special occasion, celebration, or activity that occurs at a specific time of the year, typically annually. 
These events can often associated with cultural, religious, or societal traditions and are influenced by the changing seasons. 
 
Seasonal events are of high importance as they can potentially lead to delays in ranking systems in adopting more seasonal ads into impressions.
As seasonal ads have a short lifetime, this delay can cause a shift in the behavior of ranking models, further dragging seasonal ads into impressions when the seasonal event has been passed, further degrading performance.
Hence, the ability to \emph{proactively} detect seasonal ads using multimodal content, provides a solution to remedy this issue.

%% file: sections/gt_collection.tex
In this section, we discuss the challenges of data collection and annotation, and detail our proposed ground truth collection pipeline as an effective best practice:
In Section~\ref{lab:keywork_pipeline}, we explain how we collected our initial datapoints for various events by leveraging a keyword-based method that can collect relevant data with high precision.
In Section~\ref{lab:human_labeling}, we describe the human labelling pipeline we leveraged in conducting this research.
And finally, in Section~\ref{lab:mm_llm_annot}, we provide a solution that we have explored during this research exploration, that leverages a multimodal LLM in the annotation and ground-truth collection loop.

\subsection{Keyword filtering}
\label{lab:keywork_pipeline}

\begin{figure}[h]
  \centering
  \includegraphics[width=1\linewidth]{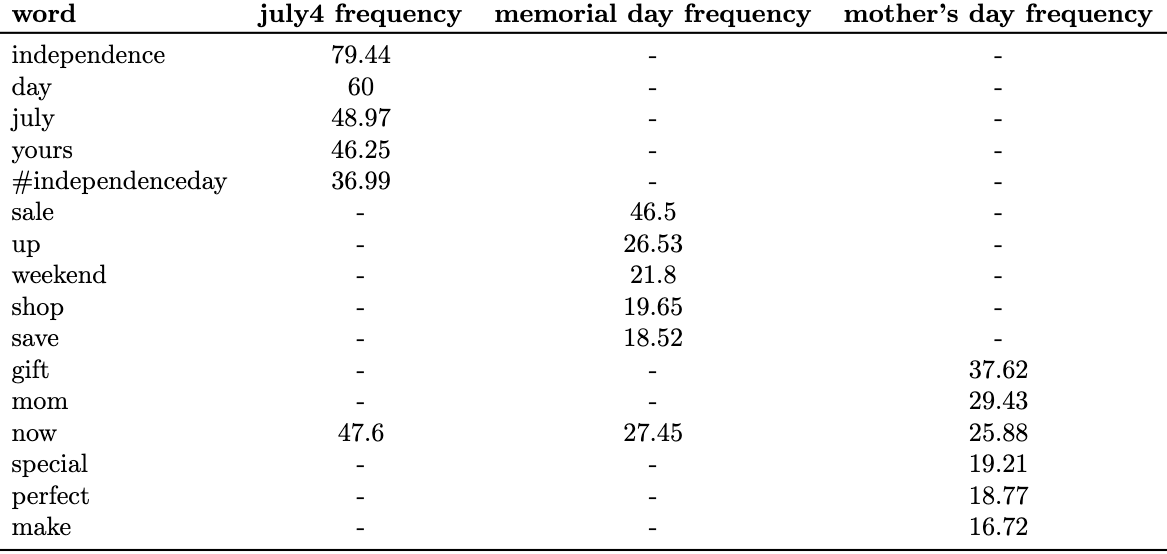}
  \caption{Top secondary keywords observed in seasonal ads during May}
\label{tab:secondary_keyword}
\end{figure}

Traditionally, the use of relevant keywords to gather data from a specific category has been a key factor in industrial systems due to its scalability, cost-effectiveness, and quick turnaround time.
Hence, a straight forward solution to identify ads related to a particular event is by leveraging the keywords associated exclusively with the event. 
For the US market for instance, these keywords can be “Easter”, “Thanksgiving”, “Memorial day“. 
Given a large corpus of ads, this method can bring a considerable amount of ads with a high precision (98\% in our estimations). We call this high precision set the \emph{primary dataset}.
Though we estimate the coverage of this method to be around 10\%. 
The next step to increase the coverage for an event using a keyword-based approach, is to find more keywords associated primarily with that event, which we denote as the \emph{secondary dataset}. 
To detect the secondary keywords, we used the primary unique keywords to select event-related subsets of ads. Figure~\ref{tab:secondary_keyword} shows an example of the keywords detected based on top frequency for 3 different events. 
A set of \emph{secondary keywords} that are not directly synonymous, but are strongly relevant, have been used to achieve additional coverage. The recall of this method has estimated to be around 30\% with a significantly lower precision: approximately 10\%. As a point of comparison, we estimated a set of random ads having around 2\% precision.
Thus we use the primary dataset (with 98\% precision) for benchmarking of detection methods, and recommend as the best practice for building benchmarks.
The secondary dataset though can help us to get the coverage required for measurements of the performance of seasonal ads, which can be iteratively improved by the predictions of the detection models or human labeling.

\subsection{Human Labeling}
\label{lab:human_labeling}

We consider human labeling as an alternative method to keyword filtering, to achieve additional coverage in the areas not fully supported by the other methods such as keyword-based selection.
Deciding whether a specific ad is related to a given event can be very subjective. 
For instance, a discount offer ending on February the 14th may or may not be related to Valentine's day. 
Hence, we expect the precision of human labeling to be lower than the high precision keyword-based approach and with a significant increase in recall.

\begin{figure}[!th]
\centering
    \includegraphics[width=1\linewidth]{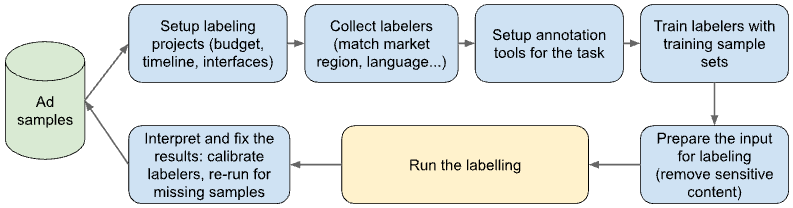}
    \caption{Human labelling pipeline for seasonal ads}
\label{lab:labeling}
\end{figure}

To study this, we ran a human labeling experiment based on 2000 samples through a crowd sourced human labeling service. We detail the process in Figure~\ref{lab:labeling}.
The performance of the labeling in the experiment resulted in the following: precision: 60\%, recall: 89\%, F1: 72\%.
Hence, we conclude that labeling with crowdsourced annotators proved to be useful for ground truth collection, but relatively expensive and time consuming. 
We additionally observed that some steps, such as filtering ad content before labeling, can increase the amount of required work, and may introduce delays, which may further reduce the benefits of outsourcing the labeling.

The blue boxes on the Figure~\ref{lab:labeling} are the additional steps in the human labeling pipeline, as compared to other approaches such as keyword filtering or leveraging LLMs, which will get into in Sections~\ref{lab:keywork_pipeline} and \ref{lab:mm_llm_annot}, respectively.

\subsection{Leveraging Multimodal LLMs}
\label{lab:mm_llm_annot}

Multimodal Large Language Models (MLMs) have emerged as cutting-edge tools for understanding content, capable of grasping subtle elements such as humor or emotion in images. These models process and integrate various data types—including text, images, and audio or video—to surpass the capabilities of traditional machine learning models.\\
\textbf{Alignment and Fusion}: MLMs merge input embeddings from different sources into a cohesive representation. For example, text from an ad and images from its thumbnail are processed through separate encoders, and their outputs are combined. Techniques like cross-modal attention mechanisms can be used to underline interactions between text and images, focusing on specific features from each modality that enhance the understanding of seasonality in advertisements. This method capitalizes on the strengths of each modality, capturing subtleties that might be overlooked when using them separately. See~\ref{subsection:mmllm-model} for more details.\\
\textbf{Finetuning and Optimization}: We have developed an MLM approach tailored to detect seasonal cues by finetuning it with a dataset of labeled seasonal advertisements. 
This process involves finetuning the LLM weights to enable it to accurately identify ads based on their seasonality.

Furthermore, seasonal ads may be unevenly distributed across categories, which can introduce biases due to data imbalance. We address this by upsampling minority classes to counteract these biases.

%% file: sections/method.tex
\subsection{Task}
Given the multi-modal representation of an ad, such as title, texts, and images, the task is to classify whether the ad is seasonal related to one of the promotional events or holidays.

For robustness evaluations, we create two versions of train and test sets: one with the original texts in the ad, and the other with the primary keywords related to the seasonal event removed (e.g. ‘Valentine’). 
We would like the models to be able to make the call based on a thorough understanding of the whole ad content, instead of memorizing or recognizing keywords only.

\subsection{Detection Models}
We leverage two models for this task: a) an end-to-end MLM, and b) late-fused MLM.
While our e2e MLM leverages an encoder-decoder architecture and is more capable in terms of expressively, prior multimodal knowledge, and reasoning capabilities, the late-fuse MLM uses an encoder-only MLM, which brings faster inference and better scalability tradeoffs.

\subsubsection{End-to-end MLM}\label{subsection:mmllm-model}
\textbf{Backend}: We study the utilization of LLAVA as the MLM backbone for our study. 
LLAVA connects a pre-trained vision encoder and a LLAMA-based LLM by leveraging a projection matrix, which projects the image embeddings generated by the vision encoder to the language space.
An architecture diagram of our E2E MLM has been provided in appendix~\ref{sec:model_arch}.\\
\textbf{Finetuning}: On top of LLAVA, we add a linear layer to project the output embeddings to binary classes for our classification task. During finetuning, the projection matrix, language model weights, and the output projection layer weights are updated, while the vision encoder is frozen.
\textbf{Prompt Design}:
Given the text (tile, body) and image inputs for the Ad, we ask the model to determine whether the ad is seasonal. We also explain the definition of seasonality and the event, and encourage the model to think step by step. We are also actively researching the design of the prompts to yield better results.

\subsubsection{Late-fused MLM Model}
\textbf{Backend}: As an alternative to LLAVA, we utilize a discriminative encoder-only MLM, namely a CLIP model and extract both image and text embeddings. Encoder-only MLMs offer a more efficient inference, hence, better scalability compared to generative MLMs such as LLAVA.

\textbf{Finetuning}: We leverage an MLP neural network to train a seasonality detection model via the extracted embeddings.

\subsection{Benchmarking Datasets}
We use 2 sets of benchmarking datasets: 1) single-event and 2) multi-event.
The single-event dataset uses valentine's day as the target event, whereas the multi-event dataset consists of 7 events namely Easter, Father's day, July 4th, Memorial day, Mother's day, Valentine's day, and a non-seasonal category (None).
We use the single-event benchmark for a wide range of model performance analysis, and use the best performing parameters in the multi-event benchmark.
This allows a faster iteration, while avoiding overfitting.
We utilize heuristics such as keyword matching (e.g. Valentine) to generate positive samples for fine tuning. These samples usually demonstrate high precision but lower recall and volume. We randomly downsample negatives to balance the positive and negative ratio to roughly 50/50.
Meanwhile, we launched a human annotation process to collect labels for evaluation and fine-tune.

%% file: sections/experimental_setup.tex
\subsection{ Setup}
We are interested in validating or nullifying below hypotheses:
1) MLM is able to perform well on ads seasonality detection, and 2) Finetuning is essential for performance improvement compared with zero-shot.

We summarize our findings as follows (see Appendix B for details):

1) We observe that additional modality (image) leads to a better performance in both e2e and late-fused MLMs.
2) We observed that pushing the model to learn harder cases improves the model’s generalizability.
3) An overview of model performance w.r.t finetune volume for the e2e model can be found in Figure~\ref{fig:model_perf_ablation} which is a useful guide for efficiency.
4) While e2e offers a better performance overall in all cases studied, late-fused MLM is competitive, striking a balance between computational efficiency and performance, offering a solution for large-scale industrial applications.

\subsection{Single-event performance}
Evaluated by F1 score, we see in Figure~\ref{fig:results}.a, that the best performing models are able to reach >0.95 for both precision and recall, resulting in an overall good performance of 0.98 F1 score. Among all, we find the best performing model is the one fine-tuned on more than 100K data, with both images and texts as input, and on a harder task of removing keywords in the finetune set.

\subsection{Multievent performance}
\begin{figure}
\centering
\begin{subfigure}{0.44\textwidth}
    \centering
    \includegraphics[width=\textwidth]{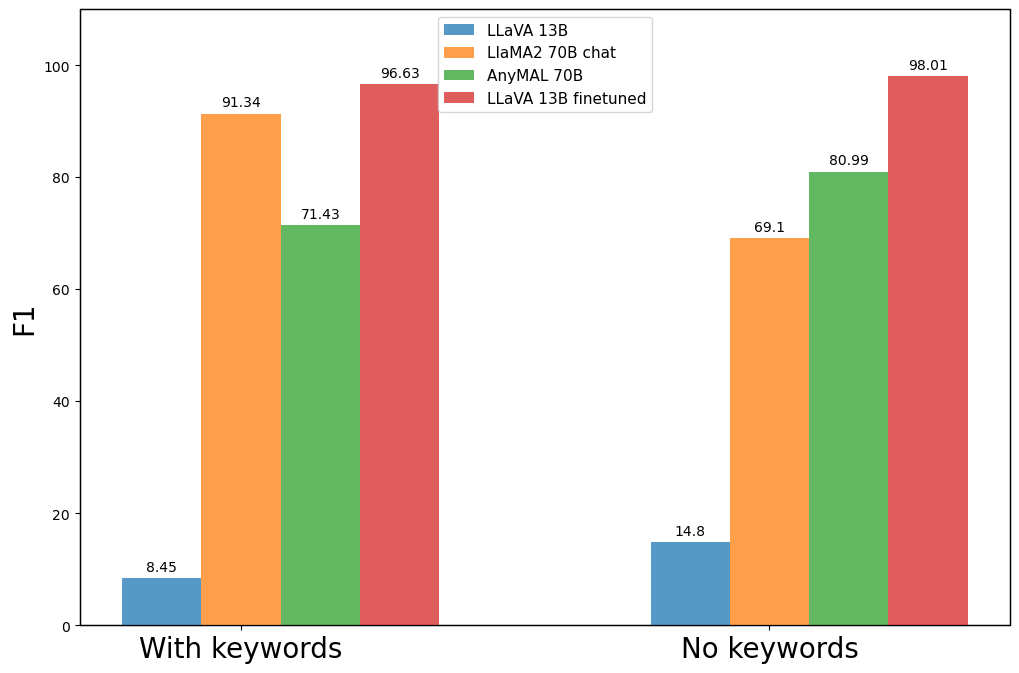}
    \caption{Model comparison (e2e MLM)}
    \label{fig:comparison}
\end{subfigure}\hfill

\begin{subfigure}{0.44\textwidth}
    \centering
    \includegraphics[width=\textwidth]{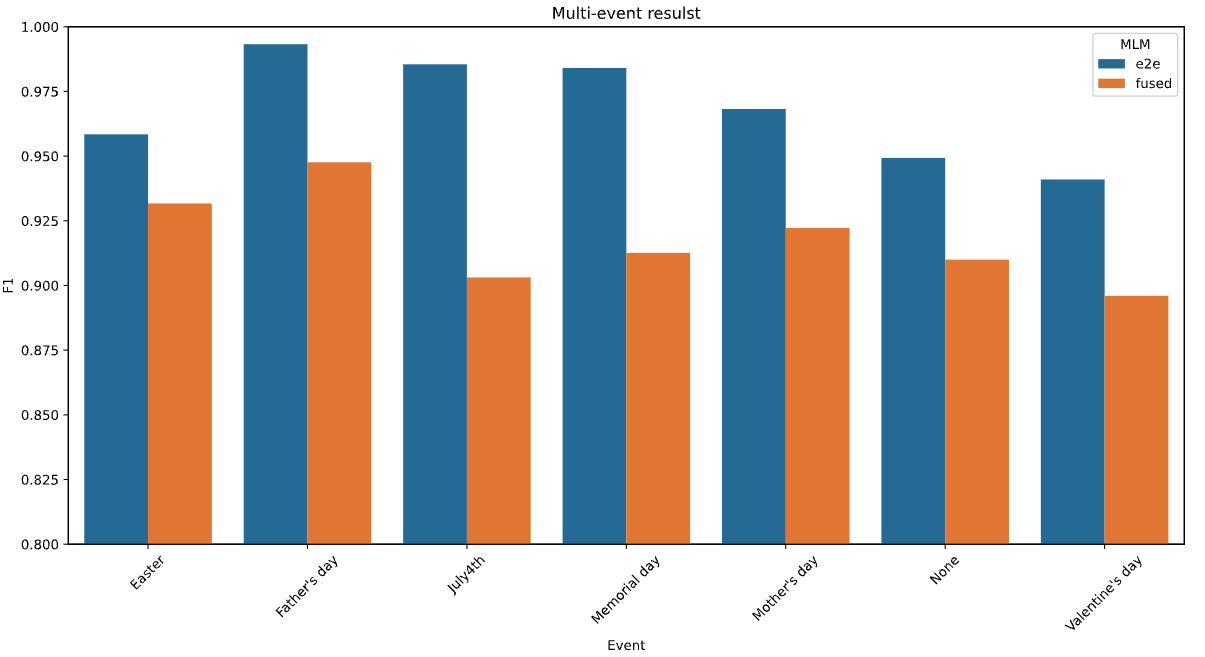}
    \caption{Multi-event results}
    \label{fig:multievent_f1}
\end{subfigure}
\caption{
a) An overview of model performance, evaluated by F1 score for single-event e2e MLM. 
b) Multievent results
}
\label{fig:results}
\end{figure}

This section expands previous single-event results to a 7-event multi-event detection results namely Easter, Father’s day, July 4th, Memorial day, Mother’s day,
Valentine’s day, and a non-seasonal category (None).
As shown in Figure~\ref{fig:results}.b, the e2e MLM achieves an averaged-F1 of ~0.97, while the late-fused results in a 0.91 averaged F1 across all events.
We present the multi-event-based F1 in Figure~\ref{fig:multievent_f1}.
As can be seen, the e2e offers high F1 across events, offer a better option for annotation purposes.
The late-fused MLM also demonstrates competitive results with a reasonable performance while offering faster inference and lower latency, and less memory requirement, hence, a better candidate for large-scale usecases in industrial ranking systems.
We additionally provide extensive ablations in our experiments, which are included in appendix~\ref{sec:ablations} due to space constraints.

\subsubsection{Comparison of backends for e2e MLM: LLAMA2 VS. AnyMAL}
In Figure~\ref{fig:model_perf_ablation} in the appendix~\ref{sec:ablations}, we cross-compared the performance of LLAVA-13B with LLAMA2-70B-Chat~\cite{touvron2023llama} and AnyMAL~\cite{moon2023anymal} and. For both AnyMAL and LLAMA2, we made inferences by calling model APIs without attempting to finetune the model.
As can be seen from the plot, the finetuned version of LLAVA-13B outperformed both models with size 70B on both datasets, with or without keywords, resulting in the best overall performance on Ads seasonality detection task.
The poor zero-shot performance of untuned LLAVA-13B also justifies the necessity of fine-tuning.

We also observed that both LLAMA2 and AnyMAL’s output could be unnecessarily loquacious as they were not tailored towards the classification task, but when successful, could provide informative reasoning steps.

%% file: sections/related_work.tex
\subsection{Seasonality detection problem}
Traditionally, detecting and identifying seasonal trends have been identified empirically as a part of time series detection and forecasting family of algorithms. 
Some of the prominent examples of this include the STL algorithm \cite{stl} which is a method for decomposing a time series into its trend, seasonal, and residual components. 
Other algorithms such as SARIMA \cite{sarima_model} (Seasonal Autoregressive Integrated Moving Average), involves modeling seasonality to capture complex seasonal patterns in time series data. These algorithms have been used in several applications ranging from medical~\cite{sarima_health_1,sarima_health_2}, sales predictions~\cite{sarima_sales} and stock value prediction~\cite{sarima_stock}. 

In these settings, seasonality is viewed purely from a perspective of composition of a single signal, while the nature of the setting that we are presenting involves other signals (such as multimodal content), which can be used to detect seasonality. In a content-based approach, seasonality is detected rather detected from the content of the ad, and not using the target signal (engagements with the ad). 
The following subsection details the related literature of multimodal content models and detecting concepts that are close to seasonality, though, to the best of our knowledge, our work is the first of its kind that leverages multimodal content for detecting seasonal events within the scope of seasonality detection for ads delivery systems. 

\subsection{Multimodal Large Language Models for concept detection}
Recent advancements in the field of natural language processing have witnessed the proliferation of large language models (LLMs) as a preeminent technique for facilitating content understanding within extensive systems, particularly in applications such as search engines, recommendation platforms, advertising platforms and scenarios involving vast repositories of information.

Early efforts~\cite{universal_sentence_encoder,bert} in developing general-purpose language models consists of deep learning models capable of encoding sentences into low-dimensional vectors to facilitate efficient text-based retrieval and semantic similarity calculations. 
BERT is one good example that has significantly improved the state-of-the-art for numerous natural language processing tasks such as text understanding and retrieval.

Subsequently, the emergence of large language models (LLMs) has further revolutionized the field. LLMs with few shot learning capabilities~\cite{few_shot_learners} have demonstrated that large language models such as LLAMA series~\cite{touvron2023llama} or GPT-3~\cite{few_shot_learners} can perform few-shot learning, thereby enhancing content understanding and generation across diverse domains. Further Lewis et al.~\cite{aug_generation} showcased the potential of these models in addressing knowledge-intensive natural language processing tasks by effectively integrating retrieval mechanisms with generation processes.

The advent of multimodal large language models (MM-LLMs) has significantly augmented the capabilities of traditional LLMs, helping them to integrate visual information into undertanding of their content. Consequently, these models are increasingly demonstrating improved performance in various applications involving multimodal content analysis and representation.

Prominent MM-LLMs such as CLIP~\cite{clip}, Flamingo~\cite{flamingo}, LLaVA~\cite{llava} , BLIP-2~\cite{blip2} integrate images and help in enhanced content understanding to better tasks such as search and recommendation through finetuning. 
Often, this involves fusing the text and image embeddings \cite{CommerceMM,imageBind} and some systems \cite{meta_transformer} have demonstrated an ability to fuse up to 12 different modalities. 
In this work, we aim to leverage finetuned MultiModal LLMs and their enhanced capacity for conceptual understanding as a means of generating labels pertinent to the task of seasonality detection

\subsection{Seasonality in downstream delivery systems}
Studying the effects of seasonality to improve delivery systems has mainly been studied from the lens of search and recommendation. 
Stormer el al.\cite{stormer_et_al} proposes an algorithm to improve recommender systems by identifying seasonal products and filtering recommendations based on the current season. The work suggests automatically detecting the high and low seasons for products based on transaction data, and only recommending seasonal products during their high season. The seasonality detection part of the work is done purely through sales numbers and no context understanding is employed.
Volpe et al.\cite{volpe_et_al}  describe integration of seasonality for recommender systems in two-stage ranking models. Following the retrieval stage, the ranking stage employs a Gradient Boosting model to select the final recommendations, incorporating seasonality information through features like seasonal tendency which captures the portion of purchases for a product in each season. Ma et al.\cite{ma_et_al} perform similar studies in the effects of seasonality in the Grocery reommendation. Kramar et al. \cite{kramar_et_al} discuss seasonality as a novel context for web search, suggesting that user behavior and interests may vary based on seasonality, similar to how it impacts product relevance in recommender systems. However, in all these works, content based seasonal understanding isn't a prominent part of the work and seasonality detection is still done empirically.
Yang et al. \cite{Yang2021} propose product attribute based seasonality detection in their work. This is then used to incorporate seasonality into their downstream search system. 

To the best of our knowledge, learning seasonality as a concept from content through MM-LLMs, and via leveraging contextual seasonal understanding, have not been explored before for the purpose of proactive detection as a usecase in ads delivery downstream.